\documentclass[superscriptaddress, twocolumn, showpacs, showkeys]{revtex4}%
\usepackage{hyperref}%
\usepackage{amsfonts}%
\usepackage{amssymb}%
\usepackage{bm}%
\usepackage{graphicx}%
\begin{document}
\title{Direct observation of non-local effects in a superconductor}
\date{\today}
\author{A. Suter}%
\email{andreas.suter@psi.ch}%
\affiliation{Laboratory for Muon Spin Spectroscopy, PSI, CH-5232 Villigen PSI, Switzerland}%
\author{E. Morenzoni}%
\email{elvezio.morenzoni@psi.ch}%
\affiliation{Laboratory for Muon Spin Spectroscopy, PSI, CH-5232 Villigen PSI, Switzerland}%
\author{R. Khasanov}%
\affiliation{Laboratory for Muon Spin Spectroscopy, PSI, CH-5232 Villigen PSI, Switzerland}%
\affiliation{Physics Institute, University of Zurich, CH-8057 Zurich, Switzerland}%
\author{H. Luetkens}%
\affiliation{Laboratory for Muon Spin Spectroscopy, PSI, CH-5232 Villigen PSI, Switzerland}%
\affiliation{Institut f\"{u}r Metallphysik und Nukleare
             Festk\"{o}rperphysik,  TU Braunschweig, 38106 Braunschweig, Germany}%
\author{T. Prokscha}%
\affiliation{Laboratory for Muon Spin Spectroscopy, PSI, CH-5232 Villigen PSI, Switzerland}%
\author{N. Garifianov}%
\affiliation{Kazan Physical-Technical Institute, 420029 Kazan, Russian Federation}%
\begin{abstract}%
We have used the technique of low energy muon spin rotation to
measure the local magnetic field profile $B(z)$ beneath the
surface of a lead film maintained in the Meissner state ($z$ depth
from the surface, $z \lesssim 200$~nm). The data unambiguously
show that $B(z)$ clearly deviates from an exponential law and
represent the first direct, model independent proof for a
non-local response in a superconductor.
\end{abstract}%
\maketitle
Applying a small magnetic field parallel to the surface of a
superconductor results in the expulsion of the magnetic flux from
its interior, except for a small region on the nm scale close to
its surface where the local field $B(z)$, measured at a depth $z$
from the surface, is heavily damped (Meissner-Ochsenfeld effect).
The spatial field dependence $B(z)$ reflects the electromagnetic
response of the superconductor and yields valuable information
about its nature. If the Cooper pairs can be handled as point
like, the electrodynamics of the system can be treated as local
and the field extends exponentially over a typical length
$\lambda$ (London penetration depth) of the order of a few tens to
hundreds nm. Non-local effects in superconductors arise when the
variation of the electromagnetic field over the extent of the
pairs cannot be neglected \cite{pippard53,bcs57}. This is for
instance the case in conventional superconductors with coherence
length $\xi \gtrsim \lambda$ or at nodes of the energy gap of
unconventional superconductors, where the $k$-dependent coherence
length becomes effectively infinite \cite{kosztin97}. In
$\mathrm{YBa_2Cu_3O_{6.95}}$ indications of non-local/non-linear
effects have been found in the field dependence of the effective
magnetic penetration depth as determined by $\mu$SR measurements
in the bulk of the vortex lattice \cite{sonier99,amin00}.

In conventional superconductors the penetration profile of the
magnetic field in the Meissner state is predicted to clearly
deviate from the usual exponential decay at the surface. Though
theoretical predictions are already half a century old, the direct
experimental verification has been lacking. Signatures for
non-local field penetration have been searched so far by induction
techniques \cite{drangeid62}, magnetoabsorption resonance
spectroscopy \cite{doezema84,doezema86}, and polarized neutron
scattering reflectrometry (PNR) \cite{nutley94}. The pioneering
measurements of Drangheid and Sommerhalder \cite{drangeid62}
revealed that there is a sign reversal of $B(z)$ (as predicted by
theory), but no further quantitative results could be drawn from
this experiment. The magnetoabsorption resonance spectroscopy
technique uses the fact that quasi-particles travelling parallel
to the shielding current are bound to the surface by an effective
magnetic potential. Indication of non-local effects in Al was
inferred by comparing microwave induced resonant transitions
between the energy levels of these bound states with transition
fields calculated from the energy levels of the trapping
potential, parameterized to include the shape of the non-local
BCS-like potential \cite{doezema84,doezema86}. Due to the resonant
character of the experiment, only few specific points of the
potential are probed. In addition the normal metallic state has to
be understood very well in order to interpret the data. This,
together with uncertainties in modelling the surface bound states,
makes the confirmation of the predicted functional form of $B(z)$
rather indirect. The specular reflectivity of neutrons spin
polarized parallel or anti-parallel to $\bm{B}$ also depends on
the field profile. However, this technique requires model-fitting
of spin-dependent scattering intensities rather than giving a
direct measure of the spatial variation of the magnetic field. Up
to now non-local corrections have been found to lie beyond the
sensitivity of PNR \cite{nutley94}.

A direct measurement of $B(z)$ requires experimental probes that
allow to measure microscopically the magnetic properties of  the
region extending only a few tens of nm away from the surface. We
used the newly developed low energy muon spin rotation technique
(LE-$\mu$SR) to map $B(z)$ in superconducting lead
\cite{morenzoni94}. We find that the functional dependence of
$B(z)$ is indeed non-exponential and that it follows the predicted
Pippard-BCS theory
\cite{pippard53,bcs57,nam67a,nam67b,halbritter71}. We obtain a
value for the London penetration depth $\lambda_{\rm L} =
57(2)$~nm and a clean-limit coherence length $\xi_0 = 90(5)$~nm
for Pb.

A weak external magnetic field acts on the ground state of the
superconductor as a perturbation. Within a perturbation expansion
one can show \cite{schrieffer64,tinkham80,poole95} that the
following non-local relation between the supercurrent density
$\bm{j}$ and the vector potential $\bm{A}$ holds (in Coulomb gauge
$\nabla \cdot \bm{A} = 0$):

\begin{widetext}
\begin{equation}\label{eq:jA_nonlocal}
  j_\alpha(\bm{r}) = - \sum_\beta\, \int
    \underbrace{\left[R_{\alpha\beta}(\bm{r}-\bm{r}')-
    \frac{e^2 n_{\rm S}}{m^*}\,\delta(\bm{r}-\bm{r}')
    \delta_{\alpha\beta}
    \right]}_{\displaystyle =: K_{\alpha\beta}(\bm{r}-\bm{r}')}\,
    A_\beta(\bm{r}')\; \mathrm{d}\bm{r}'
\end{equation}
\end{widetext}

\noindent where $e$ is the electron charge, $n_{\rm S}$ the
supercarrier density, $m^*$ the effective electron mass, and
$\nabla \wedge \bm{A} = \bm{B}$. The first term in the square
brackets, $R_{\alpha\beta}$, describes the paramagnetic response,
whereas the second reflects the diamagnetic one. $K_{\alpha\beta}$
is called the kernel. If the wave function of the electronic
ground-state were ``rigid'' with respect to all perturbations
(rather than only those which lead to transverse excitations)
$R_{\alpha\beta}$ would be identically zero and
Eq.(\ref{eq:jA_nonlocal}) would reduce to the local
$\bm{j}$-$\bm{A}$ relation

\begin{equation}\label{eq:jA_local}
  j_\alpha(\bm{r}) = -\frac{1}{\mu_0 \lambda_{\rm L}^2} A_\alpha(\bm{r})
\end{equation}

\noindent with $\mu_0$ the magnetic permeability of the vacuum.
This combined with the Maxwell equation $\nabla \wedge \bm{B} =
\mu_0 \bm{j}$ yields, at a plane superconductor-vacuum interface,
the result of an exponentially suppressed magnetic field

\begin{equation}
  B(z) = B_{\rm ext}\, \exp(-z/\lambda_{\rm L})
\end{equation}

\noindent with the London penetration depth $\lambda_{\rm L} =
\sqrt{\displaystyle\frac{m^*}{\mu_0 e^2 n_{\rm S}}}$, which is the
well known result.

However, $R_{\alpha\beta}$ has a range of the order of the
diameter of the Cooper pairs, {\em i.e.} of the coherence length
$\xi$. The magnetic penetration depth sets the length scale for
the decay of the magnetization; for $\lambda \gg \xi$ the spatial
variation of the vector potential $\bm{A}$ over the
superconducting pairs is negligible and the one-parameter local
description of Eq.(\ref{eq:jA_local}) holds. If $\xi \gtrsim
\lambda$ the full non-local description has to be taken into
account. Fourier analysis of the perturbation and of the response
show that \cite{specReflec}

\begin{equation}\label{eq:Bz_nonlocal}
  B(z) = B_{\rm ext} \int \frac{q}{q^2 + \mu_0 K(q\xi, T, \ell)}\,
         \sin(q z)\; \mathrm{d}q
\end{equation}

\noindent where $q$ is the wave vector, $T$ the temperature, and
$\ell$ the electron mean free path. The functional form of
$K(q\xi, T, \ell)$ (Fourier transform of the kernel in
Eq.(\ref{eq:jA_nonlocal})), starting from microscopic
considerations, is explicitly known in the weak (BCS)
\cite{bcs57,halbritter71} and the strong coupling limit
\cite{nam67a,nam67b,swihart71}. The resulting formulae are rather
involved but are very close to the phenomenological expression of
Pippard \cite{tinkham80,pippard53}

\begin{equation}\label{eq:kernel}
  \mu_0 K(x,T,\ell) = \frac{1}{\lambda_{\rm L}^2}\, \frac{\xi(T,\ell)}{\xi(0,\ell)}
    \left[ \frac{3}{2}\, \frac{1}{x^3} \left\{ \left( 1+x^2 \right)
    \arctan(x) - x \right\}\right]
\end{equation}

\noindent with $x = q \xi(T, \ell)$  and
$\displaystyle\frac{1}{\xi(T, \ell)} = \frac{J(0,T)}{\xi_0} +
\frac{1}{\ell}$, where $\xi_0$  is the clean limit coherence
length and $J(0,T)$ is given according to Ref. \cite{bcs57}. The
kernel has the property $\mu_0 K(x\to 0, T\to 0) = \lambda_{\rm
L}^{-2}$ (corresponding to the local limit). This holds for the
BCS kernel as well.

\begin{figure}[h]
  \centering
  \includegraphics[width=0.55\textwidth]{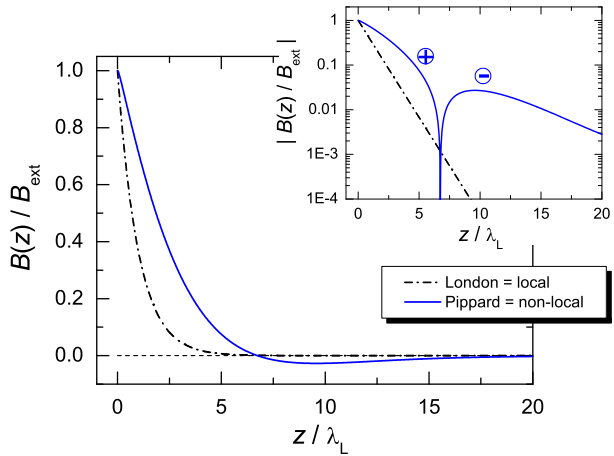}\\
  \vspace{-5mm}
  \caption{Comparison between local and non-local magnetic penetration
           profiles for the same $\lambda_{\rm L}$  (parameters from Al for
           $T\to 0$, $\lambda_{\rm L} = 18$ nm, $\xi_0 = 1600$ nm). The
           dashed-dotted line shows the typical exponential field profile
           predicted from the London theory. The solid line shows the non-local
           field profile from BCS theory. Specific features are: (i) The
           penetration profile is non exponential. (ii) The initial slope
           is less steep than in the local approximation. (iii) There is a
           field reversal before $B(z)$ decays towards zero. (iv) The inset
           shows that $\log(B(z))$  vs.\ $z$, exhibits a clear curvature.}\label{fig:Bz_theory}
\end{figure}

Fig. \ref{fig:Bz_theory} shows the theoretical predictions in the
case that $\xi\gg\lambda$. The main features can be understood
qualitatively. In the non-local case the perturbing field changes
over the extension of the Cooper pairs; since the charges within a
Cooper pair do not experience the same force, the screening
response is less effective and hence the field falls initially
less rapidly than in the case of a point response. With the field
penetrating further, at some range Cooper pairs
``overcompensate'', which accounts for the curvature of
$\log(B(z))$ as well as for the field reversal of $B(z)$
\cite{drangeid62}.

We have used the 100\% polarized low energy muon beam at the Paul
Scherrer Institute and the muon spin rotation technique ($\mu$SR)
to directly determine the values of the magnetic field as a
function of depth underneath the surface \cite{morenzoni94}. With
a tunable energy between 0.5 and 30 keV these particles are
implanted one at a time at variable depth between $\sim 1$~nm and
a few hundreds nm beneath the surface of the specimen. The local
magnetic field $B(z)$ at the stop position causes the muon spin to
precess. The temporal evolution of the spin polarization of the
muon ensemble, $P(t)$, is monitored by the detection of the decay
positrons which are anisotropically emitted preferentially in the
direction of the muon spin at the moment of the decay
\cite{blundell99}. This quantity is directly related by a Fourier
transform to the internal magnetic field distribution sensed by
the muon ensemble. The field distribution $p(B)$ is connected to
the implantation profile of the muons $n(z,E)$ by:

\begin{equation}\label{eq:nz_pB}
  n(z,E)\; \mathrm{d}z = p(B)\; \mathrm{d}B
\end{equation}

\noindent which states that the probability that a muon will
experience a field in the interval $[B, B + \mathrm{d}B]$ is given
by the probability that it will stop at a depth in the range $[z,
z + \mathrm{d}z]$. Integrating Eq. (\ref{eq:nz_pB}) on both sides
yields:

\begin{equation}\label{eq:nz_pB_int}
  \int_0^z n(\zeta, E) \mathrm{d}\zeta = \int_B^\infty p(\beta)
     \mathrm{d}\beta,
\end{equation}

\noindent which, for a chosen $z$, is an equation for $B$. We
calculate $n(z,E)$  with the Monte Carlo code {\tt TRIM.SP}
\cite{eckstein91}, which yields reliable implantation profiles for
the muons, as shown in Ref.\ \cite{morenzoni02}. Since $n(z,E)$ is
known and $p(B)$ is measured, $B(z)$ can be uniquely determined.

To search for non-local effects we investigated thin films of
type-I superconducting Pb. The films were sputtered directly onto
sapphire or quartz crystals mounted on a He flow cryostat. The
samples had a diameter of 50 mm and thickness of 1055(50) nm
(sample I) and 430(20) nm (sample II) as determined by a high
sensitivity surface profiler and Rutherford backscattering (RBS).
For sample I an oxide-layer of 5.8(3) nm was found by RBS, whereas
sample II had an oxide-layer of 16(2) nm. The critical temperature
was determined by means of resistivity and susceptibility
measurements to $T_{\rm c} = 7.21(1)$~K. The mean free path   was
estimated to be $\simeq 100$~nm from resistivity measurements.
After zero field cooling a $B_{\rm ext} = 8.82(6)$ mT was applied
parallel to the surface. In the experiment low energy muons,
$\mu^+$, with their spin perpendicular to the magnetic field and
their momentum were implanted in the samples at variable depth up
to 150 nm and the decay positrons detected by scintillation
counters surrounding the sample. Details and characteristics of
the low energy $\mu^+$ source and spectrometer are given in
\cite{morenzoni00}. The distribution $p(B)$ was derived by maximum
entropy Fourier analysis of the muon spin precession frequency of
the decay positron histograms
\cite{skilling84,rainford94,riseman00}.

\begin{figure}[t]
  \centering
  \includegraphics[width=0.55\textwidth]{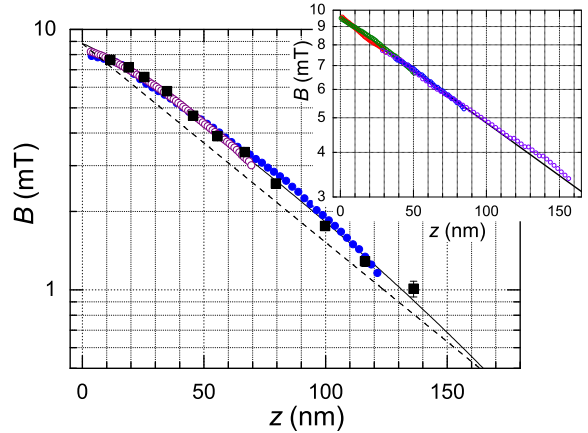}\\
  \vspace{-5mm}
  \caption{Measured magnetic penetration profile $B(z)$ of Pb in the
           Meissner state for $T = 3.05$~K [$T_{\rm c} = 7.21(2)$~K]
           in an external field of $8.82(6)$~mT, applied parallel to
           the surface of the film. Implantation energies $E$ of the
           muons: 5.2~keV open circles, 14.8~keV closed circles. The
           mean values are plotted as closed squares with implantation
           energies (2.5, 4.0, 5.4, 7.5, 10, 12.4, 15, 18, 22.5, 26, 30) keV
           from left to right. The solid curve is
           a fit according to Eq.(\ref{eq:Bz_nonlocal}) with the BCS parameters
           $\lambda_{\rm L} = 57(2)$~nm, $\xi_0 = 90(5)$~nm, and $\ell = 100$~nm
           (fixed). The dashed curve shows the exponential dependence
           assuming the values from the full BCS fit. The inset displays
           results of Meissner state measurements of optimally doped
           $\mathrm{YBa_2Cu_3O_{7-\delta}}$ at $T = 20$~K ($T_{\rm c} = 87.5$~K).
           This curve is purely exponential, as expected.}\label{fig:Pb305}
\end{figure}

\begin{figure}[h]
  \centering
  \includegraphics[width=0.5\textwidth]{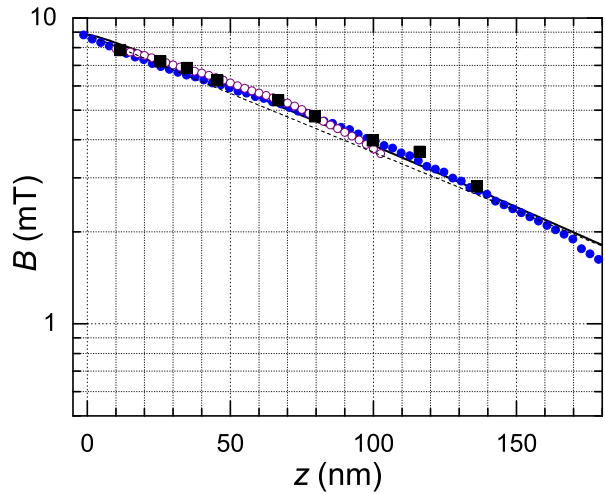}\\
  \vspace{-5mm}
  \caption{Magnetic penetration profile $B(z)$ of Pb in the Meissner state
           for $T = 6.66$~K ($T_{\rm c} = 7.21(2)$~K) in an external field
           of 8.82(6) mT, applied parallel to the surface of the film.
           Implantation energies of the muons: 14.8 keV open circles,
           30.0 keV closed circles. The mean values are shown as closed
           squares with implantation energies (2.5, 5.4, 7.5, 10, 15, 18,
           22.5, 26, 30) keV. The solid curve is a fit according to BCS theory,
           whereas the dashed curve shows an exponential with the same
           $\lambda_{\rm L}$.}\label{fig:Pb666}
\end{figure}

The measured magnetic field penetration profiles $B(z)$ in the
Meissner state of Pb at T = 3.05(3) K and T = 6.66(3) K,
respectively, are shown in the Fig. \ref{fig:Pb305} and
\ref{fig:Pb666} (sample I). Sample II gives consistent results.
The curves show clear deviations from the exponential behavior
with the characteristic curvature.  Qualitatively, the initial
slope of the curve is determined by $\lambda$, whereas the
$\log(B(z))$ curvature is mainly governed by the ratio
$\xi/\lambda$. The deviations are more pronounced at lower
temperature: on approaching $T_{\rm c}$ the superconductor becomes
more and more local. This is a consequence of the fact that
$\lambda$ has a pronounced temperature dependence close to $T_{\rm
c}$, but $\xi$ has not. The reversal of the penetrating field
could not be detected since the muon range in the experiment was
not large enough. We would like to point out that from Eq.
(\ref{eq:nz_pB_int}) it follows that the functional relationship
$B(z)$ can de determined from overlapping implantation profiles
$n(z,E)$ obtained at different energies. In order not to overload
the figures only some energies are shown (displayed in different
colors). Fig. 2 shows for instance curves obtained with 5.2 and
14.8 keV data. Their small relative deviation is a measure of the
uncertainty in the derivation of $B(z)$ and is an important
selfconsistency test of the reliability of the results. For a more
detailed discussion and a compilation of additional data we refer
to a forthcoming publication. Figs. \ref{fig:Pb305} and
\ref{fig:Pb666} also present the results from an alternative
analysis using the mean values for $\langle B \rangle = \int B
p(B)\, \mathrm{d}B$ and $\langle z \rangle = \int z n(z, E)\,
\mathrm{d}z$. The points $\langle B \rangle$ plotted versus
$\langle z \rangle$  are in very good agreement with the curve
$B(z)$ obtained from Eq.(\ref{eq:nz_pB_int}). As a further cross
check for the sensitivity and accuracy of the method based on
Eq.(\ref{eq:nz_pB_int}) we reanalyzed profiles in the Meissner
state of optimally doped $\mathrm{YBa_2Cu_3O_{7-\delta}}$ at $T =
20$ K ($T_{\rm c} = 87.5$~K, \cite{jackson00}). In this extreme
type-II superconductor, at this temperature, the data perfectly
follow an exponential law as expected for the local electrodynamic
response (Fig.\ref{fig:Pb305} inset) and are also in very good
agreement with the results previously obtained by an iterative
solution of Eq.(\ref{eq:nz_pB}). Analyzing the Pb data with the
BCS and Pippard theory leads to consistent results for
$\lambda_{\rm L}$ and $\xi_0$. The strong electron-phonon coupling
in Pb can be accounted for by a renormalization of the weak
coupling parameters of the form $\lambda_{\rm BCS} \to
\lambda_{\rm L}/\sqrt{Z}$ and $\xi_{\rm BCS} \to \xi_0 Z$ with $Z
= 1+\lambda_{\rm e-ph}\approx 2.55$ for Pb \cite{carbotte90}
($\lambda_{\rm e-ph}$, the electron phonon coupling). Compiling
all the data, we find a magnetic penetration depth $\lambda_{\rm
L} = 57(2)$~nm and a coherence length $\xi_0 = 90(5)$~nm for $T =
0$~K.

In conclusion, by using spin polarized muons of a few keV energy
as surface sensitive magnetic microprobes, we have shown that the
magnetic penetration profile at the surface of superconducting Pb
is non-exponential and that non-local electrodynamics effects, as
predicted by Pippard and BCS theory, are responsible for this
behavior. We believe that the ability to measure magnetic profiles
beneath surfaces and buried interfaces on the nm scale, by means
of LE-$\mu$SR, opens the door to explore interesting systems from
fundamental as well as from applied point of view. We look forward
to further measurements of magnetic fields and fluctuations in
surface superconductivity ({\em e.g.} proximity effects in
multilayers, surface sheath) and magnetism.

This work was performed at the Swiss Muon Source (S$\mu$S), Paul
Scherrer Institute, Villigen, Switzerland. We thank M.~Horisberger
for the preparation of the samples, M.~Doebeli for the RBS
measurements, D.~Eshchenko and D.~Ucko for the help during part of
the measurements.

\bibliographystyle{apsrev}
\bibliography{non_local}

\end{document}